# What's Wrong With Aim-Oriented Empiricism?

Nicholas Maxwell
Emeritus Reader at University College London

**Abstract**

For four decades it has been argued that we need to adopt a new conception of science called *aim-oriented empiricism*. This has far-reaching implications and repercussions for science, the philosophy of science, academic inquiry in general, the conception of rationality, and how we go about attempting to make progress towards as good a world as possible. Despite these far-reaching repercussions, aim-oriented empiricism has so far received scant attention from philosophers of science. Here, sixteen objections to the validity of the argument for aim-oriented empiricism are subjected to critical scrutiny.

**Introduction**

For four decades I have argued, in and out of print, that we need to adopt a new conception of science which I have called *aim-oriented empiricism*.[1] The argument in support of aim-oriented empiricism (AOE) seems to me to be decisive. Furthermore, as I have also argued, AOE has many enormously important implications and repercussions for science,[2] for the philosophy of science and the relationship between the two,[3] for how we conceive of rationality,[4] for academic inquiry as a whole,[5] for the task of creating a better world,[5] and indeed for any worthwhile human endeavour with problematic aims. Outstanding problems in the philosophy of science are readily solved once AOE is accepted: problems of induction,[6] verisimilitude,[7] theoretical unity[8] and the nature of scientific method.[9]

Despite all this, philosophers of science have paid scant attention to AOE.[10] This could be because the philosophy of science community is too blinkered, too prejudiced, to consider the very radical implications of the arguments for AOE, demanding as these arguments do a revolution in the way we think about science, and academic inquiry more generally. But it is, perhaps, more plausible to suppose that the argument for AOE is fatally flawed, and AOE and its implications have been ignored for so long for that reason.[11]

In this paper I propose to subject the argument for AOE to as devastating a destructive attack as I can muster. First, I state the argument; then, I do my utmost to destroy it. If I fail, I call upon others to finish off the demolition job for me – if it can be done.

**Some Implications and Repercussions of Accepting Aim-Oriented Empiricism**

What, in a little more detail, are the consequences of accepting aim-oriented empiricism? There is, to begin with, a major extension in scientific knowledge and understanding, in that, granted aim-oriented empiricism (AOE), the thesis that the universe is physically comprehensible becomes as secure an item of theoretical knowledge as anything can be in physics – more secure than our best theories, such as quantum theory and general relativity. Furthermore, what it means to assert that the universe is physically comprehensible is precisely explicated. AOE provides a rational, if non-mechanical and fallible, method for the discovery of fundamental new theories in physics. Outstanding fundamental problems in the philosophy of science are solved if AOE is accepted: the problem of induction, the problem of verisimilitude, the problem of explicating precisely what it means to say of a physical theory that it is simple, unified or explanatory, the problem of justifying persistent preference in physics for simple, unified or explanatory theories, and the problem of specifying the precise nature of scientific method in a way which does justice both to the unity and diversity of methods in science.

Instead of methods being fixed, AOE holds that the aims and methods of sciences evolve with evolving knowledge. There is something like positive feedback between improving knowledge, and improving knowledge-about-how-to-improve-knowledge. Metaphysics becomes an integral part of physics. AOE transforms both the philosophy of science and science, the philosophy of science becoming an integral part of science itself to form the new unified enterprise of natural philosophy.[12] When generalized, AOE does justice to the point that there are, not just problematic metaphysical assumptions inherent in the aims of science but, in addition, problematic assumptions concerning values, and concerning the human use of science. This generalized version of AOE provides a framework within which the humanitarian or social aims of science may be improved as science proceeds – the outcome being a kind of science much more sensitively responsive to human need.[13]

Once AOE is accepted, it becomes clear that we need a new conception of rationality designed to help us achieve what is of value whatever we may be doing, and especially when aims are problematic, as they often are. This new conception of rationality – *aim-oriented rationality* (AOR) as I have called it – arrived at by generalizing AOE, is designed to help us make progress and improve our aims and methods in life whatever we may be doing, but especially when we pursue worthwhile but problematic aims. There is the hope that, as a result of putting AOR into practice in life, we may be able to get into life – into politics, industry, commerce, agriculture, the media, the law, education, international relations – something of the astonishing progressive success achieved in science. AOR has revolutionary implications when applied to academia. It emerges that the proper basic task of social inquiry and the humanities is, not to improve knowledge and understanding of the social world, but rather to promote increasingly cooperatively rational tackling of problems of living, and to help humanity adapt our institutions and ways of life so that AOR is put into practice – especially when our basic aims and ideals are problematic. Academia as a whole comes to have, as its basic task, to seek and promote, not just knowledge, but rather wisdom – wisdom being the capacity to realize (apprehend and create) what is of value in life, for oneself and others, wisdom thus including knowledge, technological know-how and understanding, but much else besides.

The Enlightenment programme of learning from scientific progress how to achieve social progress towards an enlightened world is transformed. AOR requires that we attend to aims, values, ideals, feelings and desires as well as to fact, evidence, truth, and valid argument. Both science and art have vital rational roles to play in inquiry devoted to the pursuit of wisdom. AOR produces a synthesis of traditional rationalism and romanticism, and improves on both. AOR requires that we attend to feelings and desires to discover what is of value, but not everything that feels good is good, and not everything that we desire is desirable. We need to put mind and heart into touch with one another, so that we may develop mindful hearts and heartfelt minds.[14]

AOR, in short, has dramatic implications for universities, for academic inquiry, for our intellectual goals in doing science and the humanities. AOR has even more dramatic implications when applied to individual, social, institutional and global life. Above all, it has revolutionary implications for the effort to make progress towards as good a world as possible – an aim inherently and profoundly problematic. It is hardly too much to say that almost all our current global problems have arisen because of our long-standing failure to put AOR into practice in science, academia, politics, industry, agriculture, and many other aspects of social life. All our current global problems – climate change, rapid population growth, destruction of natural habitats and rapid extinction of species, spread of modern armaments, conventional and nuclear, the lethal character of modern warfare, vast inequalities of wealth and power around the globe – have been created by the successful pursuit of highly problematic aims which have not been subjected to the sustained imaginative, critical, and effective scrutiny that AOR would require. Humanity is in deep trouble, and we urgently need to put into practice the new kind of thinking and living that AOE and AOR require.[15]

If the argument designed to establish AOE is valid, this would seem to provide a powerful case for the wealth of intellectual, social and moral repercussions that I have just indicated. Not all of these repercussions follow logically from AOE. It is rather that the argument in support of AOE, if valid, initiates a powerful line of thought in support of these repercussions. But if the argument for AOE is not valid, much of the force of this line of thought is lost. Arguments in support of AOR, in support of scientific and academic reform, and in support of reform of our political, industrial and economic world can be developed independently of AOE, but these arguments become more cogent and forceful if AOE is indeed the correct way to think of the progress-achieving methods of natural science.

Thus, it really is a matter of some importance to determine whether the argument for AOE is valid.

**The Argument Against Standard and for Aim-Oriented Empiricism**

Most scientists and philosophers of science hold that the basic intellectual aim of science is factual truth, nothing being permanently presupposed about the truth, the basic method being to assess claims to knowledge impartially with respect to evidence. Simplicity, unity or explanatory power may influence choice of theory too, but not in such a way that the universe, or the phenomena, are assumed to be simple, unified or comprehensible. *No thesis about the world can be accepted as a part of scientific knowledge independent of evidence, let alone in violation of evidence*.

This orthodox view I call *standard empiricism* (SE). It is a tenet of such widely diverse views in the philosophy of science as logical positivism, logical empiricism, conventionalism, inductivism, hypothetico-deductivism, constructive empiricism, most versions of scientific realism, the views of Popper, Kuhn and Lakatos, and most contemporary philosophers of science  The following argument, however, seems to show that SE is untenable.

In physics, only *unified* fundamental physical theories are accepted, even though endlessly many empirically more successful disunified rival theories exist and can easily be formulated. This persistent acceptance of unified theories, even though endlessly many empirically more successful rivals can be concocted, means that physics persistently accepts, in an unacknowledged, implicit fashion, a substantial untestable (i.e. metaphysical) thesis about the nature of the universe, to the effect, at least, that it is such that no precise, seriously disunified theory is true.[16] This suffices to establish that SE is

false.[17]

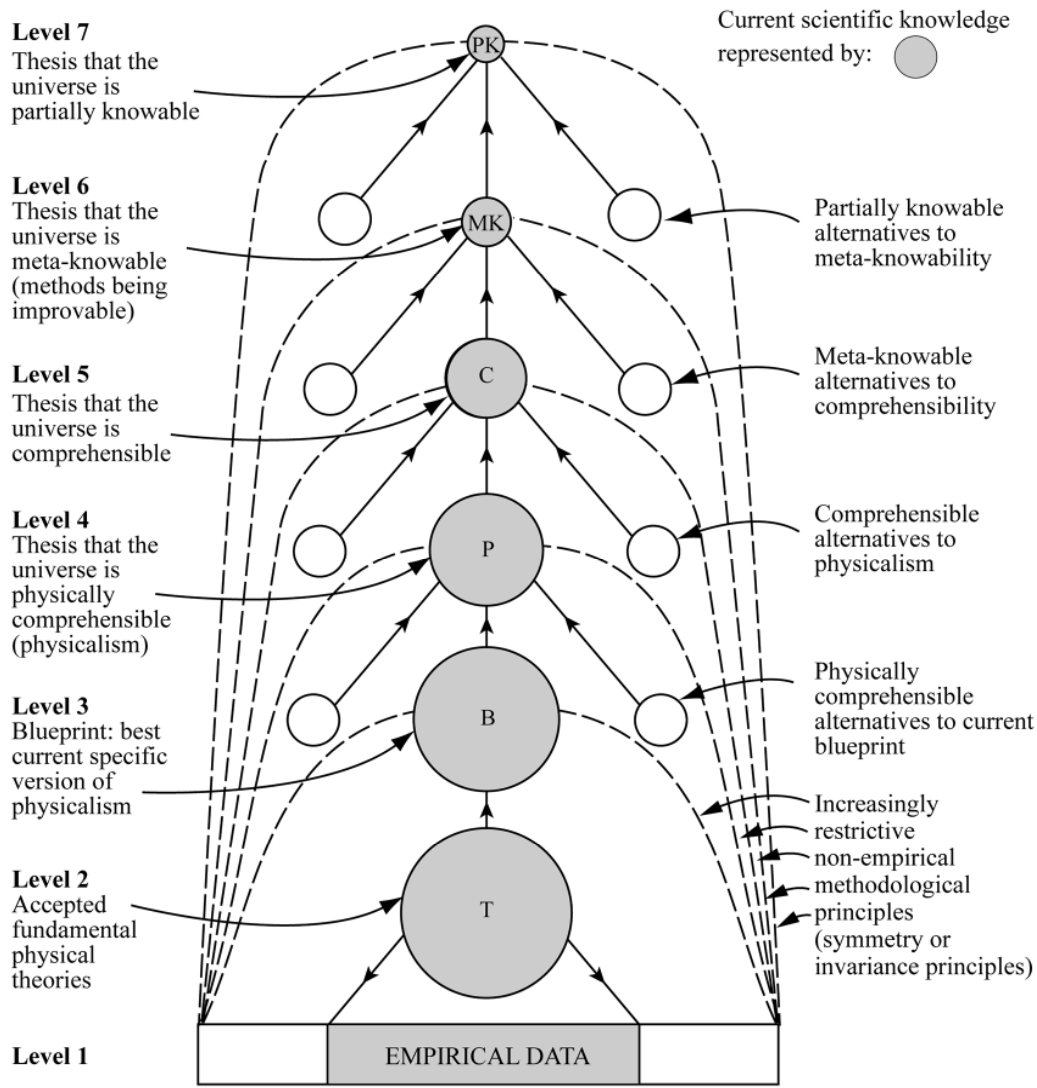

**Figure 1: Aim-Oriented Empiricism**

Once it is accepted that physics does persistently, if implicitly, accept, as a part of scientific knowledge, that there is an underlying dynamic unity in nature, to the extent, at least, that no seriously disunified theory is true, two crucial questions come to the fore:

(a) What ought this assumption to be?
(b) How can we best set about improving the current accepted version of this assumption, in the light of empirical fruitfulness and other relevant considerations?

My proposed solution to these two problems is *aim-oriented empiricism* (AOE), depicted in figure 1. The basic idea of AOE is to represent the metaphysical assumption of physics – implicit in the persistent acceptance of unified theories only, even when empirically more successful disunified rivals can be concocted – in the form of a hierarchy of assumptions concerning the comprehensibility and knowability of the universe. These assumptions assert less and less as one goes up the hierarchy, and thus become increasingly likely to be true; and they become more and more nearly such that their truth is required for science, or the pursuit of knowledge, to be possible at all. In this way a framework of relatively insubstantial, unproblematic, fixed assumptions and associated methods is created within which much more substantial and problematic assumptions and associated methods can be changed, and indeed improved, as scientific knowledge improves. Put another way, a framework of relatively unspecific, unproblematic, fixed *aims* and methods is created within which much more specific and problematic aims and methods evolve as scientific knowledge evolves. (Science has the aim, at each level, from 7 to 3, to discover in what precise way the relevant assumption is true, assumptions implicit in aims becoming increasingly substantial and problematic as one descends from level 7 to level 3.) At any level, from 6 to 3, that assumption is accepted which (a) accords best with assumptions above in the hierarchy, and (b) is associated with the most empirically progressive research programme, or holds out the greatest promise of stimulating such a programme. There is thus something like positive feedback between improving knowledge, and improving aims-and-methods, improving knowledge-about-how-to-improve-knowledge. This is the nub of scientific rationality, the methodological key to the unprecedented success of science.[18] Science adapts its nature to what it discovers about the nature of the universe (see Maxwell, 1974, 1976, 1984, 1998, 2004, 2005).

At level 7, there is the assumption that the universe is such that we can acquire some knowledge of our local circumstances. If this minimal assumption is false, we have had it whatever we assume. It can never be in our interests to abandon this assumption. At level 6 we have the more substantial and risky assumption that the universe is such that we can learn how to improve methods for improving knowledge. This promises to be too fruitful for progress in knowledge not to be accepted. At level 5 there is the assumption that the universe is comprehensible in some way or other – it being such that something exists which provides in principle one kind of explanation for all phenomena. At level 4 there is the even more substantial assumption that the universe is *physically* comprehensible, there being some kind of invariant physical entity, pervading all phenomena which (together with instantaneous states of affairs) determines (perhaps probabilistically) how events unfold in space and time. The universe is such, in other words, that the true physical "theory of everything" is unified,[19] or physically comprehensible. At level 3 there is the even more substantial assumption that the universe is physically comprehensible in some more or less specific way. Superstring theory, or M-theory, might be this assumption today. At level 2 we have currently accepted fundamental theories of physics: at present, the standard model, and general relativity. At level 1 we have accepted empirical data – low level experimental laws.

The argument for AOE has two stages. First, there is the argument that persistent acceptance of unified theories means physics makes a persistent, substantial, highly problematic and implicit metaphysical assumption about the universe. Second, there is the argument that AOE provides us with the best methodological framework for the progressive improvement of the metaphysical assumptions of physics. The first stage is crucial. If it is valid, it becomes obvious that AOE is superior to any version of SE, even if AOE may be capable of being further improved.  For, granted the validity of the first stage, AOE conforms to, and exemplifies, the following principle of intellectual rigour whereas SE violates it:-

**Pinciple of Intellectual Rigour:** An assumption that is substantial, influential, and implicit must be made explicit so that it can be critically assessed, so that alternatives can be developed and assessed, in an attempt to improve the assumption that is made.

   AOE observes this principle.  Indeed, the different levels of AOE amount to an elaboration of the principle.  But SE violates it.  This suffices to establish that AOE is to be preferred to any version of SE.

   The crucial question becomes: is the first stage of the argument for AOE valid?  It is this question that I now examine.

**Possible Objections to Argument that Science Makes a Metaphysical Assumption**
   Here are 16 objections to the argument that persistent acceptance of unified theories means that physics makes a substantial metaphysical assumption about the universe – the first stage of the argument for AOE.

   1. It is not the case that, given an accepted physical theory, endlessly many empirically more successful disunified rivals can be concocted.
   2. Even if this is the case, this does not mean that a metaphysical assumption is accepted.
   3. Preference is given in physics for unified theories because physicists seek explanatory theories: this preference does not mean a metaphysical assumption is made.
   4. Unity plays a role in the context of discovery, but not in the context of justification.
   5.  What it means to say of a theory that it is "unified" is much too vague for persistent preference for "unified" theories in physics to imply that a persistent substantial assumption about the nature of the universe, concerning "unity", is thereby made.
   6. The history of physics does not reveal the discovery of increasingly unified theories.
   7. The history of physics may reveal the persistent discovery and acceptance of increasingly unified theories, but this does not mean physics gives *a priori* preference to unified theories.
   8. If Nature had been disunified, physics would have had to accept disunified theories.  Hence physics cannot assume, and does not assume, that nature is unified.
   9. Evidence reveals that the assumption of unity meets with success.  The assumption of unity is based on evidence, and is not *a priori*.
   10. Nature might be knowable, to some extent at least, but not unified.  It might be only partially unified, or such that some other clue makes discovery of knowledge possible.
   11. Given two theories, $T_1$ and $T_2$, $T_1$ unified, $T_2$ disunified, ostensibly equally well supported by the evidence then, other things being equal, $T_1$ is actually better confirmed, better supported by the evidence.  Persistence acceptance of unified theories does not, thus, imply that a metaphysical thesis concerning unity is implicitly accepted.
   12. Evidence only supports that part of a theory that (a) predicts the evidence, and (b) can be extended to other phenomena in a unified way.  This means disunified theories ostensibly better supported by evidence than unified theories are actually not better supported.
   13. The argument against SE and for AOE is justificational in character.

14. A conjunction of two or more theories is not itself a theory.

15. At most, the argument establishes that physics assumes that nature *behaves as if* unified. It does not establish that nature *is* unified.

16. If most physicists take standard empiricism for granted, and standard empiricism is, nevertheless, untenable, and would, if implemented in scientific practice, bring physics to a standstill, how is it that physics has, despite this, met with such (apparent) success over the last three or four centuries?

I take these sixteen objections in turn.

1. It is not the case that, given an accepted physical theory, endlessly many empirically more successful disunified rivals can be concocted.

**Reply**: Consider any accepted fundamental physical theory, T (Newtonian theory, classical electrodynamics, quantum theory, general relativity, QED, or the standard model). We can readily concoct as many equally empirically successful but disunified rival theories as we please by modifying T for some as yet untested prediction only. Let us suppose that T is Newtonian theory (NT). One rival might assert: (i) everything occurs as NT predicts up until the end of 2050; after that date an inverse cube law of gravitation obtains. Another rival might assert: (ii) everything occurs as NT predicts except for stones dropped into the centre of four concentric gold rings of radius 8, 10, 13 and 20 feet placed on some flat, horizontal surface, in which case an inverse cube law obtains. Another might assert: (iii) everything occurs as NT predicts except for a system of gold spheres, each of mass greater than 1,000 tons, confined to a spherical region of outer space of 500 miles across, in which case an inverse quartic law of gravitation obtains between the spheres. In each case, (i) to (iii), endlessly many different modifications of NT can be made. Furthermore, endlessly many different untested consequences of NT can be specified. Some of these disunified or "aberrant" versions of NT can easily be refuted, but endlessly many more that are as yet unrefuted can easily be concocted. Some will presumably never be refuted in the entire history of the cosmos, for example (iii) above.

Disunified rivals that are even more empirically successful than the accepted theory, T, can be concocted as follows. T, we may suppose, (a) successfully predicts phenomena A, (b) fails to predict phenomena B because the equations of T cannot be solved, (c) is ostensibly refuted by phenomena C, and (d) fails to predict phenomena D, because these lie outside the range of predictions of T, and are not predicted by any other accepted fundamental physical theory. All we need to do to concoct an empirically more successful rival to T is modify T so that it asserts: for phenomena A, everything occurs as T predicts, and for B, C and D everything occurs as the empirically established laws assert that it does. The new theory, T*, (a) recovers the empirical success of T in A, (b) successfully predicts phenomena B that T fails to predict, (c) successfully predicts C that ostensibly refuted T, and (d) successfully predicts phenomena D that lie beyond the scope of T. T* recovers all the empirical success of T, is not refuted where T is, and predicts phenomena that T fails to predict, some of which lie beyond the scope of T. Quite properly, T* would not be considered for a moment in scientific practice because of its grossly ad hoc, disunified character. Nevertheless, T* satisfies all the requirements one could stipulate for being an empirically more successful theory than T.

It may be objected that not all accepted physical theories run into empirical difficulties, but this objection hardly stands. Much scientific research of the kind Kuhn called "normal science" is concerned to eliminate clashes between theory and empirical results. It often turns out that ostensible refutations are not real refutations: when experiments are repeated, neglected factors are taken into account, or better approximate derivations made, the clash is transformed into a successful prediction for the theory. But until this happens, the theory is ostensibly refuted by the phenomena, and T* successfully predicts phenomena that, ostensibly, refute T.[20]

One also has to take into account that experiments in physics are often difficult to perform correctly: they often initially yield result which are rejected precisely because they clash with the predictions of accepted theory. In other words, it frequently happens that initial results ostensible refute accepted theories – and the results are rejected precisely for that reason. If physics gave equal weight to unified and disunified theories, experiments would persistently favour the latter. Persistent bias towards unity affects, not just what theories are accepted but, even more startlingly, what experimental results are accepted.[21] Even at the experimental level, nature persistently says "everything is very complicated" and physicists (quite correctly) refuse to listen.

But even if we ignore (c) above, and accept the fiction that accepted theories run into no empirical difficulties anywhere in their range of applications, (b) and (d) above still provide the means to concoct empirically more successful disunified rivals to accepted theories. Furthermore, we can concoct endlessly many such empirically more successful rivals to T by first employing the strategies indicated in (i), (ii) and (iii) above, and then employing the strategies (a), (b) and (d).[22]

   2. Even if there are endlessly many empirically more successful disunified rivals to accepted theories, this does not mean that a metaphysical assumption is accepted.

   **Reply**: But why not? Suppose scientists only accepted theories that postulate atoms, and persistently rejected theories that postulate different basic physical entities, such as fields — even though many field theories can easily be, and have been, formulated which are even more empirically successful than the atomic theories — the implication would surely be quite clear. Scientists would in effect be assuming that the world is made up of atoms, all other possibilities being ruled out. The atomic assumption would be built into the way the scientific community accepts and rejects theories — built into the implicit *methods* of the community, methods which include: reject all theories that postulate entities other than atoms, whatever their empirical success might be. The scientific community would accept the assumption: the universe is such that no non-atomic theory is true.

   How does this differ in principle from the analogous situation of physics only ever accepting unified theories, even though empirically more successful disunified rivals are available? There appear to be no grounds for holding that a big assumption is implicitly being made in the first case, but no such assumption is being made in the second one.

   3. Preference is given in physics for unified theories because physicists seek explanatory theories: this preference does not mean a metaphysical assumption is made.[23]

**Reply**: As long as physics seeks truth – even if only at the empirical level – then persistent acceptance of unified theories when endlessly many empirically more successful disunified rivals are available must mean physics thereby assumes that the universe is such that all such disunified theories are false. Conceivably, it could be argued, however, that in some purely theoretical contexts, without practical applications, physics might value *explanation*, and therefore *unity*,[24] more highly than *truth*. Even though T* is empirically more successful than T, nevertheless T is accepted because of its explanatory character, even though it is known not to be such a good candidate for truth as T*. Granted these circumstances, preference for unified theories does not commit physics to assuming that disunified theories are false.[25]

But unified theories – and laws – are accepted in preference to more empirically successful disunified rivals in contexts where truth really does matter, because of practical applications such as bridge building or the design of aeroplanes. In designing and building a bridge, engineers require that physical laws about such matters as the strength of steel yield correct predictions. But, given any such accepted law, L, any number of empirically more successful disunified rivals can always be concocted which predict that the bridge will collapse. In accepting L for the purposes of building the bridge, all these empirically more successful, disunified rivals are assumed to be false. Lives are at stake. If one of them is true, the bridge will collapse and lives may well be lost.

A basic task of physics is to provide laws and theories which are such that their predictions can be relied on for practical purposes. Here, physics seeks to accept laws and theories that are true – to the extent, at least, that the standard empirical predictions are true. In persistently rejecting empirically more successful disunified rivals, physics does in effect assume that these rivals are all false – which amounts to assuming that the universe is such that disunified theories are false whatever their empirical success may be.

If persistent acceptance of unified theories (when endlessly many empirically more successful disunified rivals are available) only occurred when the aim is to improve theoretical explanation and understanding for their own sake, the objection under consideration might have some plausibility. For, we might value the explanatory character of a theory so highly that we are prepared to overlook its inadequacies as far as truthfulness is concerned. But unified laws and theories are accepted (in preference to empirically more successful disunified ones) when what is accepted will be used for practical purposes – building bridges or designing aeroplanes – where truth (of predictions, at least) is all important. In these widespread contexts, physics cannot be justified in sacrificing truth for explanation. In a court of law, it would not be convincing to argue, when a bridge collapses, killing a number of people, that the theory used to design the bridge was employed because of its explanatory character, a rival theory, more likely to be true, being rejected because it was not explanatory. To sacrifice truth for unity, in such widespread contexts, would be equivalent to being prepared to sacrifice *human lives* for unity.[26]

4. Unity plays a role in the context of discovery, but not in the context of justification.

**Reply**: But, as I have argued above, unity plays a vital role when it comes to *acceptance* of laws and theories in physics, even for practical applications, and thus in the so-called context of justification.

5. What it means to say of a theory that it is "unified" is much too vague for persistent preference for "unified" theories in physics to imply that a persistent substantial assumption about the nature of the universe, concerning "unity", is thereby made.

**Reply**: A number of inadequate attempts have been made to solve the problem of what it means to say of a physical theory that it is "unified".[27] Even Einstein recognized the problem but confessed that he did not know how to solve it.[28] The problem is, however, readily solved within the framework of AOE: see works referred to in note 8. Here I shall be brief. One major difficulty that confronts the attempt to specify what it is for a theory to be "unified" stems from the fact that the simplest, most unified theory can be reformulated to assume as complex, disunified a form as one could wish, and *vice versa*. In order to overcome this difficulty, the crucial step one needs to take is to appreciate that "unity" applies, not to the theory itself, its form, axiomatic structure or pattern of inferences, but to *what the theory asserts about the world* – to the *content* of the theory, in other words. Immediately, variability of formulation is irrelevant. As long as all the different formulations of a theory *have the same content*, the degree of unity of the theory is the same.

For unity, we require of a theory that it has the same content throughout all the actual and possible phenomena to which the theory applies. The theory must assert that the same dynamical laws govern the evolution of phenomena throughout all the possible phenomena specified by the theory, and to which the theory applies. Here "the same dynamical laws" means "same in content", not "same in formulation". If dynamical laws differ in N regions of the space of all possible phenomena specified by the theory, then the degree of unity of the theory = N. For unity, we require N = 1.

But now there is a refinement. Dynamical laws may differ in different ways, some ways being more serious than others. They may differ (1) in different space-time regions; (2) because one or more specially restricted, dynamically special object exists; (3) because dynamical laws change as the value of some variable changes – for example, mass; (4) because the theory postulates two or more distinct forces; (5) because the theory postulates physical entities with different dynamical properties, such as mass or charge; (6) because the theory fails to unify space-time and matter (by means of a broken symmetry, perhaps).[29] We may take these six types of unity to be accumulative, in that unity of type (4), let us suppose, requires that unity of types (3), (2) and (1) have also been achieved. For N = 1, we require that space, or space-time, and matter are unified. As long as a theory fails to achieve this, its degree of unity must be such that N ≥ 2. This acknowledges that empty space, or space-time, is a possibility, and thus a possible distinct physical entity.

The upshot is that what "unified theory" means is not vague at all; rather, there are at least six different kinds of unity, each of which may have any degree N ≥ 2 (except for the sixth kind of unity, for which we might have N = 1).

Corresponding to any particular kind of unity, (n), with n = 1,2,…6, and any particular degree of unity, N, there is a particular metaphysical thesis asserting that the universe is such that its fundamental dynamical laws are unified in kind and degree (n,N).  For perfect unity we require (n = 6, N = 1).  The existence of this infinite range of possibilities provides a part of the case for adopting AOE, so that physics has the best available tools for discovering in what precise way, and to what precise extent, the universe is unified.

In so far as physics today does not even consider, let alone accept, theories with type (1), (2) or (3) disunity to any degree N > 2, whatever the empirical success of the theory might be, if considered, it follows that physics persistently assumes that the universe is such that no theory is true that is disunified in these ways, and to these degrees.[30]

6. The history of physics does not reveal the discovery of increasingly unified theories.

**Reply**: On the contrary, this is just what the history of physics does reveal.  Every theoretical revolution in physics brings about greater unification.  Newton, in unifying Kepler and Galileo, unified terrestrial and astronomical motion – unification of type (1).  Maxwell, in unifying electricity, magnetism and optics, brought about unification of types (4) and (5).  Special relativity brought greater unity to Maxwell's classical theory of electrodynamics, unified energy and mass by means of $E = mc^2$, and partially unified space and time to form space-time – unifications of types (4) and (5).  General relativity dissolved the force of gravitation into a richer conception of space-time – unification of type (4).  It also holds out the hope of unification of type (6).  The theory of elements and chemical compounds brought astonishing unification to chemistry, in reducing millions of different sorts of elementary substances to the less than 100 of the elements.  Quantum theory and the theory of atomic structure brought massive unification to atomic theory, properties of matter, interactions between matter and light.  Instead of nearly 100 elements plus electromagnetic radiation, the theory postulates just four entities: the electron, proton, neutron and photon – unification of types (4 and (5).  Instead of a multiplicity of laws concerning the chemical and physical properties of matter, there is Schrödinger's equation – again, unification of types (4) and (5).  Quantum electrodynamics unifies quantum theory, special relativity and classical electrodynamics – unification of type (4).  The electro-weak theory of Weinberg and Salam partially unifies the electromagnetic and weak forces – unification of type (4).  The quark theory of Gell-Mann and Zweig brought greater unity to the theory of fundamental particles: a large number of hadrons were reduced to just six quarks – unification of type (5).  Quantum chromodynamics brought further unification to the theory of fundamental particles by providing a quantum theory of the strong force – unification of type (4).  The standard model, the current quantum theory of fundamental particles and the forces between them, partially unifies the electromagnetic, weak and strong force – partial unification of type (5).  The unification is only partial because the different forces are all locally gauge invariant, but different kinds of locally gauge invariant forces nevertheless, observing different symmetries.  And the theory postulates a number of distinct particles with different, even though related, properties.  Superstring theory attempts to reduce all particles to just one kind of entity – the quantum string in ten or eleven dimensions of space-time, observing just one law of evolution.  If superstring theory ever achieves a precise formulation, it would constitute unification of type (6) – the ultimate aspiration of theoretical physics.[31]

7. The history of physics may reveal the persistent discovery and acceptance of increasingly unified theories, but this does not mean physics gives *a priori* preference to unified theories.

   **Reply**: It does not, and nor is that the argument deployed above in support of the assertion that physics makes a persistent, substantial, highly problematic metaphysical assumption.

8. If Nature had been disunified, physics would have had to accept disunified theories. Hence physics cannot assume, and does not assume, that nature is unified.

   **Reply**: However disunified nature might have been, and however disunified might be the laws and theories accepted by physics as the best available, nevertheless endlessly many even more disunified could easily be concocted which would fit phenomena even better than accepted laws and theories. My reply to objection 1 establishes this point. We can imagine worlds quite different from how this one seems to be in which non-empirical constraints on what theories are to be accepted are very different from those that operate in theoretical physics today. The hierarchy of theses of AOE is intended in part to take such possibilities into account. In such worlds, metaphysical theses would be accepted different from those accepted by science today – the crucial point being that *some* kind of metaphysical thesis would be accepted, implicit in the adoption of non-empirical constraints on what theories can be accepted. What is not possible is to proceed fully in accordance with standard empiricism, the *only* consideration governing acceptance and rejection of theory being evidence (and, perhaps, empirical content).

9. Evidence reveals that the assumption of unity meets with success. The assumption of unity is based on evidence, and is not *a priori*.

   **Reply**: A basic idea of AOE is that metaphysical theses, low down in the hierarchy, are chosen in the light of (a) compatibility with theses higher up in the hierarchy, and (b) capacity to generate, or at least accord with, empirical success, at levels 2 and 1. So 9 does not constitute an objection to AOE.
   As long as physics persists in rejecting all theories that are disunified (in one or other of the precise senses of "disunified" specified in the answer to objection 5) whatever their empirical success might potentially be, physics persists in assuming that the universe is such that all disunified theories (in the relevant sense) are false. With respect to physics conducted in this way, this thesis of unity functions as an *a priori* assumption. Furthermore, as long as the assumption is unrecognised and implicit, it is dogmatically maintained. The great benefit gained from honestly acknowledging the assumption is that it becomes possible to subject this highly influential and problematic conjecture to sustained imaginative and critical scrutiny in an attempt to improve it, transform it into something closer to the truth, and even more empirically fruitful for science. The best way to do this is to embed it in the hierarchical structure of AOE.
   These considerations cannot be taken to imply, however, that there is no *a priori* element in science. If *all* the theses in the hierarchy of AOE were selected on the basis of empirical fruitfulness, the following fatal objection would arise. Given current scientific knowledge and AOE as indicated above in figure 1, we could always formulate a disunified empirically more successful rival, T* say, to accepted fundamental physical theories, T (the standard model and general relativity), and then formulate a "disunified" rival, AOE* say, to AOE, AOE* being such that T* is dramatically preferred to T. Only by adopting a genuinely *a priori* items of knowledge, at levels 7 and 6, can AOE* and T* be ruled out. For a detailed discussion of how this is to be done, see Maxwell (2007, ch. 14, section 6). It is this that solves the problem of induction.

10. Nature might be knowable, to some extent at least, but not unified. It might be only partially unified, or such that some other clue makes discovery of knowledge possible.

**Reply**: Nature might indeed be partially knowable, and only partially unified or such that some other clue makes the acquisition of knowledge possible. AOE allows for such possibilities. For a discussion of more than 20 such possibilities, see Maxwell (1998, pp. 168-172).

11. Given two theories, $T_1$ and $T_2$, $T_1$ unified, $T_2$ disunified, ostensibly equally well supported by the evidence then, other things being equal, $T_1$ is actually better confirmed, better supported by the evidence. Persistence acceptance of unified theories does not, thus, imply that a metaphysical thesis concerning unity is implicitly accepted.

**Reply**: If the universe is physically comprehensible in the sense that there is a yet-to-be-discovered true physical "theory of everything" that is unified, then the policy of regarding $T_1$ better confirmed than $T_2$ will meet with success. But in a universe appropriately disunified in some way, this policy will not meet with success. What this establishes is that any "principle of confirmation" which holds that, given two theories, $T_1$ and $T_2$, ostensibly equally well supported by evidence, nevertheless one *kind* of theory is better confirmed, will in fact only work in certain sorts of universe – universes which are such that theories of the specified *kind* are true (or, at least, capable of being empirically more successful, other things being equal, than other kinds of theory). In short, such "principles of confirmation" make implicit metaphysical assumptions. All such "principles of confirmation" are unacceptable, because they do not provide the means for critically assessing and improving the implicitly, and dogmatically, assumed metaphysics. It is just this which AOE does provide. Thus, any account of confirmation along the lines indicated in the above objection *would*, if adopted by science, commit science to making an implicit metaphysical assumption. Furthermore, science pursued in such a way is less rigorous than science pursued in accordance with AOE because, whereas the former denies the presence of the metaphysical assumption and therefore ensures that it is dogmatically adopted, held in such a way that it is immune to criticism and revision, AOE by contrast acknowledges the metaphysical assumption explicitly, and facilitates its critical assessment and improvement as an integral part of physics. AOE observes the above "principle of intellectual rigour", whereas any version of confirmation theory along the lines indicated above violates it.

In short, objection 11 must be rejected.

12. Evidence only supports that part of a theory that (a) predicts the evidence, and (b) can be extended to other phenomena in a unified way. This means disunified theories ostensibly better supported by evidence than unified theories are actually not better supported.

**Reply**: This objection fails for the same reason that objection 11 fails. Doing science in accordance with such a conception of confirmation will only meet with success in a physically comprehensible universe. In certain sorts of incomprehensible universes, it will fail, whereas methods which favour certain kinds of disunified, empirically more successful theories will succeed. What this demonstrates, yet again, is that any such account of confirmation implicitly attributes a more or less substantial metaphysical assumption to science. But this assumption is disavowed. That means it is adopted dogmatically. It is not open to explicit criticism and revision. AOE, which acknowledges the metaphysical assumption explicitly, and provides the methodological means for its progressive improvement, provides a more rigorous conception of science, and one which is likely to enable science to meet with greater success. In any case, objection 12 fails.

13. The argument against SE and for AOE is justificational in character.

**Reply**: "It is a justificationist thesis" claims Miller (2006, p. 94) of my argument for AOE. If I tried to justify the truth of the metaphysical theses I claim physics presupposes, Miller's charge would be correct. But my argument for making explicit metaphysical theses implicit in physics is the very opposite of that. It is that it is precisely because there is no justification for the truth of these metaphysical theses that we need to make them explicit within physics so that they may be critically assessed and, we may hope, improved.[32] AOE facilitates this process of improvement by concentrating criticism and the search for better alternatives where it is most likely to be fruitful, low down in the hierarchy of theses. Those theses are favoured which (a) accord best with theses higher up in the hierarchy, and (b) sustain the most empirically successful empirical research programmes at levels 1 and 2, or hold out the greatest hope of doing so. AOE is thus thoroughly Popperian in spirit, even though it differs sharply from Popper's conception of science.[33]

Even though there is no justification of the *truth* of any metaphysical thesis, there is a justification for the *acceptance* of these theses, granted that our concern is to improve our knowledge of truth. This justification for acceptance emerges from the solution to the problem of induction which AOE provides.[34]

14. A conjunction of two or more theories is not itself a theory.

**Reply**: If this is accepted then, so it would seem, accepted physical theories do not have empirically more successful disunified rivals because these rivals are all conjunctions of two or more theories, and thus not themselves theories. But the argument against SE cannot be demolished in this way, by a mere terminological convention. Let us call the empirically more successful disunified rivals to accepted theories "theoretical possibilities" (a new technical term). In persistently failing even to consider such theoretical possibilities, even in practical contexts where lives are at stake, scientists implicitly and persistently take for granted that they are all false, and thus make an implicit, persistent assumption about the nature of the universe. The validity of this argument is unaffected by the decision to refuse the title of "theory" to a theoretical possibility made up of two or more distinct theories.

15. At most, the argument establishes that physics assumes that nature *behaves as if* unified. It does not establish that nature *is* unified.

**Reply**: Even if this is accepted, a version of AOE is still established, one which holds that science accepts a hierarchy of metaphysical theses about no more than *the way observable phenomena behave*. This is a version of AOE which even Bas van Fraassen might accept.[35] There is, however, something very odd and counter-intuitive about this van Fraassen version of AOE.

In order to appreciate this point, consider the following three theories. $T_1$ is the theory of atomic structure plus quantum theory. The millions of different substances of chemistry are reduced to just three distinct entities: the electron, the proton and the neutron. Millions of distinct laws associated with the physical and chemical properties of matter, and the way matter and electromagnetic radiation interact, are reduced to Shrödinger's equation and laws specifying the physical properties of the electron, proton, neutron and photon.[36] $T_1$ asserts that electrons, protons, neutrons and photons do actually exist and matter and light is made up of them. $T_2$ asserts merely that phenomena occur *as if* electrons, protons, etc., exist, but does not assert that they do exist. $T_3$ is an empirically more successful, disunified rival of $T_1$ and $T_2$. $T_3$ asserts, we may assume, that the phenomena behave *as if* electrons, etc. exist for a wide range of phenomena, but not for phenomena X, let us suppose, which refute the existence of subatomic particles.

The crucial point is now this. If we attend to what these three theories *assert about the world*, whereas $T_1$ asserts that there is underlying order and unity in the vast multiplicity and apparent randomness of physical and chemical properties of matter, $T_2$ and $T_3$ make no such assertion. $T_1$ provides an *explanation* for the multiplicity and apparent randomness of physical and chemical properties of matter, in the sense that, if electrons etc. really do exist with the dynamical properties attributed to them by $T_1$, then all these physical and chemical properties of matter must be just as they are, or electrons etc. could not exist. Granted the existence of electrons etc., no modification of physical or chemical properties of matter is possible, however slight. There is a physical *explanation* as to why they have to be as they are. But $T_2$ and $T_3$ provide no such physical explanation. Both theories assert the existence of a multiplicity of apparently arbitrary physical and chemical properties of matter, *and nothing in addition*. The greater unity of $T_2$ over $T_3$ is purely formal: it has nothing to do with what $T_2$ and $T_3$ assert about the world (which, in both cases, is restricted to the purely empirical). In the one case all the phenomena occur *as if* electrons etc. exist, in the other case only *some* of the phenomena occur as if electrons exist. Neither theory asserts that electrons etc. really do exist. Hence, from the standpoint of what these two theories assert about the world, they are equally appallingly disunified. They merely assert the existence of a multiplicity of diverse, arbitrary laws governing the physical and chemical properties of matter. Whereas AOE provides a rationale for preferring $T_1$ to $T_3$, the van Fraassen version of AOE can provide no rationale for the corresponding preference for $T_2$ over $T_3$.[37]

16. If most physicists take standard empiricism for granted, and standard empiricism is, nevertheless, untenable, and would, if implemented in scientific practice, bring physics to a standstill, how is it that physics has, despite this, met with such (apparent) success over the last three or four centuries?

**Reply**: I have argued above that the physics community persistently accepts unified theories only, even though empirically more successful disunified rivals have always been available. What this demonstrates is that physicists do not implement standard empiricism in anything like a strict way when it comes to doing physics – and it is that which makes progress possible. The non-empirical requirements a physical theory must satisfy in order to be acceptable have evolved since Galileo's time, or Newton's, and to that extent, something like aim-oriented empiricism has been implemented in scientific practice. As I have argued elsewhere, something close to aim-oriented empiricism began to be put explicitly into scientific practice for the first time with Einstein's discovery of special and general relativity (Maxwell, 1993a, pp. 275-305). Following his lead, physicists have subsequently taken it for granted that new acceptable theories must satisfy symmetry principles, such as gauge invariance or supersymmetry and, in searching for new theories, they have sought to develop such symmetry principles. Wigner, for one, has emphasized the methodological significance of Einstein's work on relativity in this respect: see Wigner (1970, ch. 2). Science is, nevertheless, harmed by the current somewhat hypocritical allegiance of the scientific community to standard empiricism. Many benefits would accrue from scientists openly and wholeheartedly adopting and implementing aim-oriented empiricism: see Maxwell (2004, chs. 2-4; 2007a; 2008).

**Conclusion**

I conclude that none of the above attempts to salvage standard empiricism from refutation succeeds. Aim-oriented empiricism, and the line of thought it initiates, do indeed deserve more attention than they have so far received from philosophers.

**Notes**

[1] See Maxwell (1972; 1974; 1984 or 2007a, chs. 5 & 9; 1998; 1999; 2000a; 2002; 2004; 2005; 2006; 2007a, ch, 14; 2010a, ch. 5; 2011a; 2013).

[2] See especially Maxwell (1998, chs. 1, 4 & 7; 2004, chs. 1 & 2; 2008).

[3] See especially Maxwell (1984 or 2007a, ch. 5 & 9; 1993a, pp. 275-305; 1998, 1, & 2; 2002; 2004, chs. 1, 2 & appendix; 2005; 2006; 2007, ch. 14).

[4] Maxwell (1984 or 2007a, ch. 5; 2004, ch. 3).

[5] Maxwell (1976; 1980; 1984; 2000b; 1992a; 1992b; 2004; 2007a; 2007b; 2009a; 2010a; 2012a; 2012b; 2014).

[6] Maxwell (1998; ch. 5; 2004, appendix, section 6; 2006; 2007a, ch. 14, section 6).

[7] Maxwell (1998, pp. 211-217; 2007a, ch. 14, section 5 and appendix).

[8] Maxwell (1998, ch. 4; 2004, appendix, section 2; 2007a, ch. 14, section 2).

[9] See especially Maxwell (2004, ch. 2).

[10] To be fair, it has not been entirely neglected. Favourable comments are to be found in Kneller (1978, pp. 80-87 and 90-91); Longuet-Higgins (1984), Midgley (1986); Koertge (1989); Chakravartty (1999); Smart (2000); Juhl (2000); McHenry (2000); Shanks (2000); Muller (2004);Iredale (2005); McNiven (2005); Grebovicz (2006); Perovic (2007); MacIntyre (2009, p. 358). See also chapters in Barnett and Maxwell (2008) and McHenry (2009).

[11] On the other hand, even if the argument for AOE is flawed, given the profound significance of the issues at stake, one would have thought AOE deserved serious attention – if only to establish what the flaw is. The long-standing neglect of AOE might be regarded by some as little short of a scandal.

[12] For these points see works referred to in notes 1 to 9.

[13] See Maxwell (1984 or 2007a, ch. 5; 2004, pp. 51-67; 2008).

[14] For these implications and repercussions of AOE see Maxwell (1976; 1984; 1992a; 1992b; 2000b; 2004, chs. 3 and 4; 2007a; 2007b; 2008; 2009a; 2009b; 2010a; 2012a; 2012b; 2014).

[15] See works referred to in note 14.

[16] Many imprecise disunified theories will be true even in a universe that is perfectly physically comprehensible, in that the true physical "theory of everything", T, is unified. In such a universe, T implies any number of distinct, imprecise theories applicable to restricted ranges of phenomena. True disunified (but imprecise) theories can be arrived at by conjoining two or more such true distinct theories.

[17] For the history of this argument, and for much more detailed expositions of it, see works referred to in note 1.

[18] See Maxwell (1998, pp. 17-19; 2004, chs. 1 and 2).

[19] See works referred to in note 8 for the explication of what it means to say of a physical theory that it is "unified".

[20] A famous example is the ostensibly decisive experimental refutation of special relativity soon after it was proposed by Walter Kaufmann in 1906: see Pais (1982, p. 159).

[21] Gerald Holton vividly depicts this phenomenon in his account of Robert Millikan's "oil drop" experiment designed to measure the charge of the electron. Holton reveals that Millikan repeatedly threw away trials which clashed with the result he had determined was the correct one: see Holton (1978, ch. 2, especially pp. 53-72).

[22] Strategies for concocting empirically more successful disunified rivals to accepted theories are discussed in Maxwell (1974; 1998, pp. 47-54; 2004, pp. 10-11).

[23] This objection has been spelled out powerfully by Agustin Vicente: see Vicente (2010, pp. 632-636). See, however, my reply: Maxwell (2010b, pp. 673-674).

[24] In order to be *explanatory*, a physical theory must be *unified*: see works referred to in note 8,

[25] In preferring the explanatory T to the disunified but empirically more successful T*, physicists hope, presumably, that further work on T is more likely to lead to an even more empirically successful theory than would work on T*. In the long term, in other words, physicists are concerned with truth, even in contexts without practical applications, such as cosmology and astrophysics.

[26] For further discussion of this objection, see Vicente (2010, pp. 632-640), and Maxwell (2010b, pp. 667-671).

[27] See, for example, Jeffreys and Wrinch (1921), Goodman (1972), Popper (1959, ch. 7), Sober (1975), Friedman (1974), Kitcher (1981) and Watkins (1984. pp. 203-213). For decisive criticism of these attempts see Maxwell (1998, pp. 56-68).

[28] See Einstein (1949, pp. 21-25).

[29] For a more detailed discussion, see works referred to in note 8.

[30] For the solution to further problems, and discussion of further details, see works referred to in note 8.

[31] That revolutions in physics persistently bring about unification reveals a certain inadequacy in Kuhn's depiction of scientific revolutions: see Kuhn (1970). As I have remarked elsewhere, "*Far from obliterating the idea that there is a persisting theoretical idea in physics, revolutions do just the opposite in that they all themselves actually exemplify the persisting idea of underlying unity!*": see Maxwell (1998, pp. 181).

[32] For an explication of what it means to say that successive metaphysical theses, though all false, nevertheless get closer and closer to the truth and are, in this sense, progressively improved, see Maxwell (2007a, pp.396-400, 430-433, and especially p. 397).

[33] For accounts of how AOE and falsificationism differ, see Maxwell (2006; forthcoming).

[34] See Maxwell (2007a, ch. 14, section 6). For earlier attempts at solving the problem, see Maxwell (1998, ch. 5; 2004, appendix, section 6).

[35] See van Fraassen (1980).

[36] This is a serious simplification of the actual state of affairs – but not one that affects the argument. The argument requires only that this simplified version of atomic theory is a *possibility*, not that it represents actuality.

[37] It does not help to invoke the instrumentalism of orthodox quantum theory: see Maxwell (1993b; 1994; 2011b).